\documentstyle[prl,aps,psfig]{revtex}
\begin{document}
\draft
\twocolumn[\hsize\textwidth\columnwidth\hsize\csname
@twocolumnfalse\endcsname

\widetext
\title{Peierls-like transition induced by frustration in a two-dimensional 
antiferromagnet}
\author{Federico Becca and Fr\'ed\'eric Mila}
\address{
Institut de Physique Th\'eorique, Universit\'e de Lausanne, CH-1015 Lausanne, 
Switzerland\\
}
\date{\today}
\maketitle
\begin{abstract}
We show that the introduction of frustration into the spin-$1/2$
two-dimensional (2D) antiferromagnetic Heisenberg model on 
a square lattice via a
next-nearest neighbor exchange interaction can lead to a Peierls-like
transition, from a tetragonal to an orthorhombic phase, when the spins
are coupled to adiabatic phonons.
The two different orthorhombic ground states define an Ising order parameter, 
which is expected to lead to a finite temperature transition. Implications
for ${\rm Li_2VOSiO_4}$, the first realization of that model, 
will be discussed.
\end{abstract}
]

\narrowtext
Peierls transition~\cite{peierls} is well documented for quasi-one 
dimensional systems when the coupling to the lattice is taken into
account, both for metals and for magnetic
insulators.~\cite{boucher} It corresponds to a breaking of the 
translational symmetry, with a dimerization and a doubling of 
the unit cell in the case of spin chains. 
In strictly one-dimensional systems, the transition
occurs for an infinitesimal coupling of the lattice because the gain
of electronic or magnetic energy always overcomes the loss in elastic
energy. For instance, for S=1/2 chains,
the gain in the
magnetic energy by making a lattice distortion $\delta$ is 
$\Delta E_{mag} \sim \delta^{4/3}$, whereas the loss due to elastic energy
is $\Delta E_{elastic} \sim \delta^{2}$, and therefore a small but finite
lattice displacement is always stabilized in the thermodynamic 
limit.~\cite{cross}
This result is very specific to one-dimension though, and in higher
dimensions structural transitions are not expected for generic spin
models unless other degrees
of freedom play a role, like orbital degeneracy in the cooperative 
Jahn-Teller effect.~\cite{kugel}

In this Letter, we present strong evidence that, when frustration is 
present, 2D systems can behave in a very similar way, i.e., they
can undergo a lattice instability that breaks the symmetry even
for very small spin-lattice coupling. Motivated by recent results obtained
on Li$_2$VOSiO$_4$~\cite{millet,carretta} 
(see the discussion below), we concentrate on the
so-called $J_1{-}J_2$ model on a square lattice, where $J_1$ and $J_2$
stand for nearest and next-nearest neighbor antiferromagnetic exchange
integrals. In this model,
by increasing the frustration, there is a phase transition
from N\'eel order to a spin-liquid phase 
for $J_2/J_1 \simeq 0.38$.~\cite{doucot}
The nature of the disordered phase is still under debate, 
and many different possibilities 
have been proposed.~\cite{read,gelfand,zhitomirsky,caprio}
By increasing further the frustration ($J_2/J_1 \gtrsim 0.55$), 
the {\it ordering by disorder} mechanism~\cite{villain} is expected to
stabilize a collinear order, with spins ferromagnetically aligned 
either along the $x$
or the $y$ axis and antiferromagnetically aligned along the other, 
corresponding to pitch vectors ${\bf Q}=(0,\pi)$ and
${\bf Q}=(\pi,0)$, respectively. 
In the pure spin model, Chandra and collaborators~\cite{chandra} 
argued that this residual two-fold degeneracy of the 
ground state (GS) generates a finite-temperature, Ising-like phase transition.

However, like in the Jahn-Teller effect, where the orbital degeneracy 
is lifted by the electron-vibron interaction, this degeneracy might 
{\it a priori} also be lifted by a structural distortion if the system 
is coupled to the lattice. As we shall see, this is indeed the case, and
ferromagnetic and antiferromagnetic bonds acquire different lengths. 
In that case, it is {\it not} the translational symmetry that
is broken, like in the standard Peierls transition, but the rotational 
symmetry. Still the analogy is very suggestive: Here, the equivalent of 
the two dimerized structures of the chain are two orthorhombic phases with
the longer bond along the $x$ and $y$ directions respectively, 
and we will talk about a Peierls-like transition. Note that this mechanism 
is significantly different from the one generated by thermal
fluctuations~\cite{chandra}: In particular, a structural distortion will
have specific consequences for NMR, which might have already been 
observed in Li$_2$VOSiO$_4$.~\cite{carretta}

The $J_1{-}J_2$ model coupled to adiabatic phonons is defined by 
the Hamiltonian:
\begin{eqnarray}
&&{\cal H} = 
\sum_{(n.n.)} \left \{ 
J_1(d_{ij}) {\bf S}_i \cdot {\bf S}_j +
\frac{K_1}{2} \left (
\frac{\|\delta {\bf r}_i-\delta {\bf r}_j \|}{d_{ij}^{0}} \right )^2
\right \} + \nonumber \\
&&\sum_{(n.n.n.)} \left \{ 
J_2(d_{ij}) {\bf S}_i \cdot {\bf S}_j +
\frac{K_2}{2} \left (
\frac{\|\delta {\bf r}_i-\delta {\bf r}_j \|}{d_{ij}^{0}} \right )^2
\right \},
\label{hamilt}
\end{eqnarray}
where ${\bf S}_i=(S_i^x, S_i^y, S_i^z)$ is the spin-$1/2$ operator at site $i$,
$\delta {\bf r}_i$ is the displacement of atom $i$, assumed to be in the
plane, and $d_{ij}$ is the distance between atoms $i$ and $j$. 
This model depends
on the following parameters: {\it i)} The bare values of the 
positions of the atoms 
${\bf R}_{i}^{0}$, which form a square lattice; 
{\it ii)} The bare values of the exchange integrals $J_1$ and $J_2$, 
corresponding
to the values of $J_1(d)$ and $J_2(d)$ for the bare values of the distances
$d_{ij}^{0}=\|{\bf R}_{i}^{0}-{\bf R}_{j}^{0}\|$; 
{\it iii)} The elastic coupling constants $K_1$ and $K_2$. Since these coupling
constants are effective parameters - a detailed description of the lattice
dynamics would involve elastic terms between transition metal 
atoms and ligands - we have chosen the form of Eq.~(\ref{hamilt}), 
but the results are essentially the same with and elastic term
proportional to $(d_{ij} - d_{ij}^0)^{2}/(d_{ij}^0)^{2}$;
{\it iv)} The spin-lattice coupling constants $\alpha_1$ and $\alpha_2$:
In transition metal compounds, 
super-exchange theory combined with empirical dependences of hopping
integrals on distance~\cite{harrison} leads to exchange integrals 
that vary like the inverse of the distance to a certain power $\alpha$, 
with $\alpha$ in the range $6{-}14$, and for small displacements we can write:
\begin{equation}
\label{jj}
J_{1,2}(d_{ij})=J_{1,2} 
\left ( \frac{d_{ij}^0}{d_{ij}} \right )^{\alpha_{1,2}} \simeq
J_{1,2} \left ( 1 - \alpha_{1,2} \frac{\delta d_{ij}}{d_{ij}^{0}} \right ),
\end{equation}
with $\delta d_{ij} = d_{ij} - d_{ij}^0 \sim ({\bf R}_{i}^{0}-{\bf R}_{j}^{0}) 
\cdot (\delta {\bf r}_{i}-\delta {\bf r}_{j})/d_{ij}^0$. 
Notice that the Hamiltonian
is invariant under the rescaling 
$\alpha_{1,2} \rightarrow \lambda \alpha_{1,2}$, 
$K_{1,2} \rightarrow \lambda^{2} K_{1,2}$ and 
$\delta {\bf r}_{i} \rightarrow \delta {\bf r}_{i}/\lambda$. This allows
one to fix one parameter among $\alpha_1$, $\alpha_2$, $K_1$ and $K_2$.
The physical values of the magnetic couplings
are unaffected by this transformation.

\begin{figure}
\centerline{\psfig{bbllx=70pt,bblly=150pt,bburx=550pt,bbury=690pt,%
figure=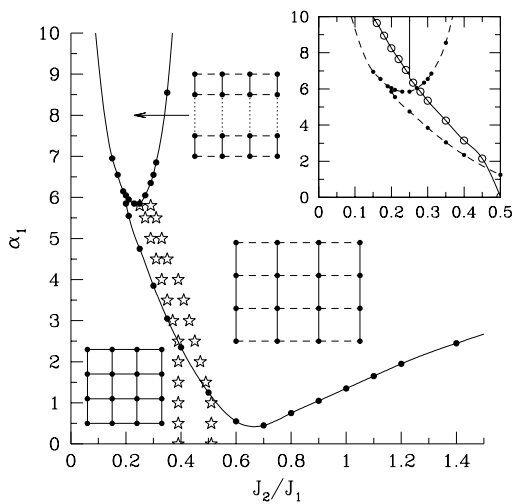,width=60mm,angle=0}}
\caption{\baselineskip .185in \label{fig1}
Structural phase diagram for the $4 \times 4$ 
cluster for $\alpha_2=\alpha_1$ and $K_2=0$. 
Points indicate the critical values for which the transition
occurs and lines are guides to the eye. Stars indicate the points with
a vanishing magnetization in
the linear spin-wave approximation, by using the optimized exchange integrals.
In each region a sketch of the lattice is depicted.
Inset: Comparison between the $4 \times 4$ (full circles and
dashed lines) and the $8$-site cluster (empty circles and solid lines).
Notice that for the $8$-site cluster the DL phase 
abruptly disappears for $J_2/J_1 > 0.25$, and that, for $J_2/J_1=0.5$,
the OL is stabilized for an infinitesimal spin-phonon coupling.}
\end{figure}

Since we are concerned with the zero-temperature properties of a 
2D system, this problem can in principle be attacked within the 
linear spin-wave (LSW) approximation. However, to perform an unbiased 
calculation requires to treat random configurations of the bond lengths,
and LSW is essentially impossible to implement in that case since
the first step, the determination of the classical GS,
has no analytic solution. So we have decided to start with 
exact diagonalization of finite clusters, which allows an unbiased 
determination of the optimal configuration, and then to use
LSW in the phases where it can provide useful insight.

To find the optimal configuration of the bond lengths
for a given value of the coupling parameters 
($\alpha_1$, $\alpha_2$, $K_1$ and $K_2$) in the subspace of $S_{tot}^z=0$,
we start from a random configuration of the site displacements
and we iteratively improve the total energy by
changing the lattice parameters $\delta d_{ij}$.
For clarity, the exchange coupling constants at equilibrium
will be denoted with tilde, e.g. $\tilde J_1$ and $\tilde J_2$, with
possibly more than two of them as we shall see.
This self-consistent Lanczos (SCL) technique
has been successfully used for different 1D and 
quasi-1D models.~\cite{poilblanc}
By working with the relative distances 
${\bf \Delta}_{ij}= \delta {\bf r}_{i}-\delta {\bf r}_{j}$, 
we are able to describe all kinds of structural displacements, including 
expansion and contraction of the lattice.

Since for a random configuration all symmetries are lost, one is limited
to rather small clusters. We have performed systematic calculations
for the $45^{\circ}$ degree tilted $8$-site cluster and 
for the $4 \times 4$ cluster, and we have checked a number of
conclusions on the $32$-site cluster. It turns out
that the $8$-site cluster is pathological for 
$J_2/J_1 >0.5$, the nearest neighbor spin-spin correlations being exactly
zero. This peculiarity introduces a very strong dependence
on the actual values of the spin-phonon coupling, which is definitely
an artifact of this lattice size. In this regard, since 
for the $4 \times 4$ cluster
the short-range correlations are finite for all finite values of $J_2/J_1$,
this cluster is expected to give more reliable insight into the
actual lattice deformations.

In Fig.~\ref{fig1}, we show the phase diagram obtained with the 
SCL method for the $4 \times 4$ cluster in the case of
$\alpha_2=\alpha_1$, $K_1/J_1=10$, and $K_2=0$.
This choice of $K_2=0$ was physically motivated by the fact that
elastic constants usually decrease very fast with distance.
However we have also determined the phase diagram for $K_2/J_1=10$.
A detailed account of the dependence of the results on the precise
value of $K_2$ and $\alpha_2$ will be given elsewhere,~\cite{becca}
but
the general conclusion is that modifications are not qualitative
but only quantitative, and quite small. Besides, thanks to the invariance
of the Hamiltonian on the rescaling mentioned above, the phase
diagram is actually the same for any ratio $K_1/J_1$ up to a
rescaling of $\alpha_1$. So, although there are a lot of parameters
in the model, the results discussed below are generic.

\begin{figure}
\centerline{\psfig{bbllx=70pt,bblly=150pt,bburx=550pt,bbury=690pt,%
figure=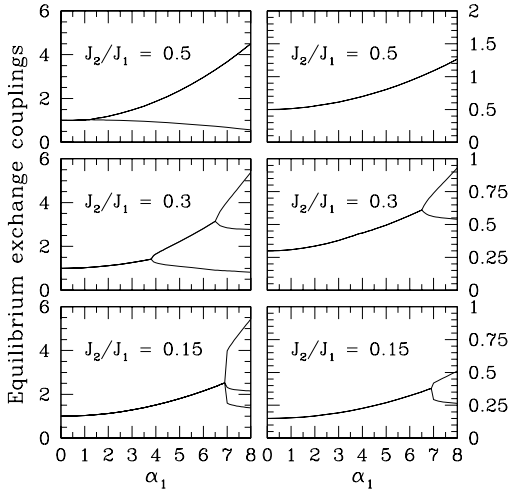,width=60mm,angle=0}}
\caption{\baselineskip .185in \label{fig2}
Equilibrium exchange integrals $\tilde J_1$ (left panels) and $\tilde J_2$ 
(right panels) as a function of the spin-phonon coupling for three different 
values of the ratio $J_2/J_1$ of the bare exchange integrals.
The parameters are those of Fig.~\ref{fig1}.}
\end{figure}

Among all the possible lattice deformations, only three are stable, 
depending on the values of the parameters: 
{\it i}) the tetragonal lattice (TL), 
with all the bond length equal, with only one value of $\tilde J_1$ 
and $\tilde J_2$. 
The TL preserves all the translational and rotational symmetries, and the
effect of the spin-phonon coupling is only to renormalize the bond
length: the GS on any finite lattice system is unique. 
{\it ii}) the orthorhombic lattice (OL), with two different
bond lengths in the $x$ and $y$ directions, inducing two different n.n.
exchange integrals
$\tilde J_1^x$ and $\tilde J_1^y$, 
but only one n.n.n. exchange integral $\tilde J_2$. 
This phase breaks the $\pi/2$ rotational symmetry and therefore
the GS is two-fold degenerate.
{\it iii}) the dimerized 
lattice (DL), with three different bond lengths, breaking the
translational symmetry of one lattice spacing in one direction, leading
to two n.n. exchange integrals in one direction, e.g., 
$\tilde J_1^y$ and $\tilde J_1^{y\prime}$,
one in the other, $\tilde J_1^x$, and two different n.n.n. exchange
integrals $\tilde J_2$ and $\tilde J_2'$.
The DL breaks both the $\pi/2$-rotation and the translational
symmetry along one axis, and the GS is four-fold degenerate.
In Fig.~\ref{fig2}, we show the behavior of the equilibrium exchange 
integrals for three different ratios of $J_2/J_1$.

The comparison between the $8$-site and the $4 \times 4$ lattice 
for $J_2/J_1<0.5$ shows that
the phase-boundaries are comparable, and that both the OL and the DL 
regions expand from 8 to 16 sites. This gives some confidence that these
phases exist in the thermodynamic limit.
As we shall see, this is confirmed for the OL by LSW. The fate of
the dimerized phase upon increasing the system size is difficult to
assess. If the $J_1{-}J_2$ model without phonons has a magnetically 
dimerized GS in the gapped phase ($0.38 \lesssim J_2/J_1 \lesssim 0.55$)
as several authors have suggested,~\cite{read,gelfand} one would 
expect this phase to expand down to zero spin-phonon coupling around
$J_2/J_1=0.5$. The results for $8$ and $16$ sites do not give any indication
of such a behavior, and larger system sizes would be necessary
to reach a definite conclusion.
Unfortunately, because of the huge amount of time needed to diagonalize
larger clusters with free bond lengths, it is impossible at present
to perform a systematic size scaling within the SCL approach.

\begin{figure}
\centerline{\psfig{bbllx=70pt,bblly=150pt,bburx=550pt,bbury=690pt,%
figure=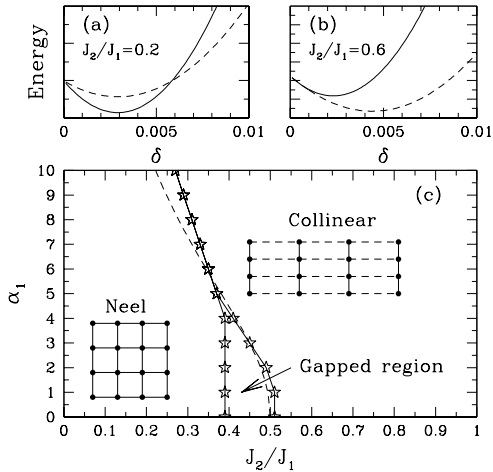,width=60mm,angle=0}}
\caption{\baselineskip .185in \label{fig3}
Energy vs lattice distortion $\delta$ within the SCSW approximation for
$K_1/J_1=10$, $K_2=0$,
$\alpha_1=0.1$, $J_2/J_1=0.2$ (a) and $J_2/J_1=0.6$ (b): The
case of $\delta_x=\delta_y=\delta$ (continuous lines) and 
$\delta_x=0$, $\delta_y=\delta$ (dashed lines) are shown.
(c): Phase diagram within the SCSW approximation for 
$K_1/J_1=10$. Stars indicate the points where
the magnetization vanishes and lines are guides to the eye. The dashed
line indicates the classical transition between the N\'eel state
and the collinear state.
In each region a sketch of the lattice is depicted.}
\end{figure}

To get further insight into the transition between TL and OL, 
we consider a similar self-consistent approach based on the spin-wave 
(SCSW) approximation of the Hamiltonian Eq.~(\ref{hamilt}).
This approximation, based on the assumption of an underlying ordered phase,
is not suitable to study the DL, which is expected to be a gapped phase.
Within the SCSW approximation, we allow the lattice to vary the bond lengths
independently along the two spatial directions, but preserving all the
translational symmetries, which only allows to describe OL and TL:
$J_1^{x,y}(\delta)=J_1(1+\alpha_1 \delta_{x,y})$ and
$J_2(\delta)=J_2[1+\alpha_1(\delta_x+\delta_y)/2]$ (here 
we only consider the case of $\alpha_2=\alpha_1$ for simplicity). 
The first step is to find the classical state.
The two local minima are the N\'eel and the collinear state (whereas
incommensurate phases are not stabilized), with energies
per site given by 
$\epsilon^{N}_{cl}(\delta)=-[J_1^{x}(\delta)+J_1^{y}(\delta)]/4
+J_2(\delta)/2+K_{1}(\delta_x^2+\delta_y^2)/2$ and
$\epsilon^{C}_{cl}(\delta)=[\pm J_1^{x}(\delta) \mp J_1^{y}(\delta)]/4
-J_2(\delta)/2+K_{1}(\delta_x^2+\delta_y^2)/2$, respectively. 
In both cases the magnetic gain is proportional to the lattice distortion. 
The stable configuration is obtained by minimizing with respect to
$\delta_x$ and $\delta_y$.
For the collinear state we find that the minimal configuration is 
obtained for $\delta_x$ different from $\delta_y$,
${\tilde \delta}_x=\alpha_1(J_2 \mp J_1)/(4K_1)$ and 
${\tilde \delta}_y=\alpha_1(J_2 \pm J_1)/(4K_1)$, and its energy per site is
${\tilde \epsilon}^{C}_{cl}=-J_2/2-\alpha_1^2(J_1^2+J_2^2)/(16K_1)$.
On the contrary, for the N\'eel state the configuration with 
${\tilde \delta}_x={\tilde \delta}_y=\alpha_1(J_1-J_2)/(4K_1)$ 
is stabilized, with 
${\tilde \epsilon}^{N}_{cl}=-(J_1-J_2)/2-\alpha_1^2(J_1-J_2)^2/(16K_1)$.
Therefore, the transition from the N\'eel to the collinear state occurs for
$J_2=J_1/[2+\alpha_1^2J_1/(4K_1)]$. As a consequence,
for $J_2/J_1 > 0.5$ an infinitesimal coupling with the 
lattice stabilizes the collinear phase, inducing a structural transition 
to an orthorhombic phase.
On the contrary, for $J_2/J_1 < 0.5$, a finite $\alpha_1$ is needed 
to obtain the orthorhombic phase.
The fact that the transition line bends towards smaller values of $J_2/J_1$ 
(see Fig.~\ref{fig3}) is just a consequence of the fact that 
$\tilde J_2/\tilde J_1 >J_2/J_1$.

The presence of the structural transitions are preserved upon introducing
quantum fluctuations at the LSW level, although the equilibrium 
distances are slightly changed from the the ones found in the 
classical limit.
The only qualitative difference is that quantum fluctuations may destroy
the magnetization near the classical transition, leading to a spin-liquid
state: The boundaries of this gapped phase are obtained
by calculating the magnetization within LSW approximation for the equilibrium 
exchange integrals.
Rather interestingly, by increasing the spin-phonon coupling, 
the disordered region shrinks,
eventually leaving the possibility of a direct transition between N\'eel and 
collinear phases.
The results of the SCSW calculation are shown in Fig.~\ref{fig3}.

Outside the spin-liquid region, there is a good quantitative agreement 
between the SCL and the SCSW calculations in
describing the TL to OL transition for small to intermediate frustration, 
whereas there is a difference for large frustration, i.e., 
$J_2/J_1 \gtrsim 0.5$.
Indeed, within the SCL method on the $4 \times 4$ cluster, the TL
structure is always stable for small spin-phonon coupling,
and this is probably an artifact of the small cluster considered.
By contrast, within the SCSW method, in the collinear phase, the lattice is
unstable toward an orthorhombic deformation for an infinitesimal spin-phonon
coupling. This is expected to be definitely the real situation, because of the
collinear spin arrangement, where the different magnetic correlations along
the $x$ and $y$ directions naturally induces an instability in the
underlying lattice structure.

From an experimental point of view, the first prototype of the
spin-$1/2$ 2D $J_1{-}J_2$ model has been recently synthetized.~\cite{millet}
It is a layered vanadium oxide ${\rm Li_2VOSiO_4}$ 
in which ${\rm VO_5}$ pyramids
are arranged in such a way that second vanadium neighbors are in the same
plane while first neighbors are not, so that $J_2$ need not be
smaller than $J_1$. There is no general agreement on the precise value 
of the ratio $J_2/J_1$, but both experimental and theoretical estimates 
lead to a ratio significantly larger than 1/2: 
NMR, muon spin rotation, magnetization and specific-heat 
measurements~\cite{carretta} suggest
that $J_2/J_1 \simeq 1.1$ with $J_1+J_2 \simeq 8 K$, while and LDA calculation
by Rosner {\it et al.}~\cite{rosner} found that $J_2/J_1$ is of the order 10.
In any case, the system is expected to develop collinear order, which
has been confirmed by NMR results at the ${\rm Li}$ site. 
Interestingly enough, the $^{29}{\rm Si}$ NMR spectrum is very 
anomalous at low temperature,~\cite{carretta} and given the very 
symmetric position of ${\rm Si}$ in the lattice, this cannot be attributed
to the development of collinear order but must be related to a 
structural phase transition. Given the rather low characteristic temperatures 
in that system ($T_c\simeq 2.8 K$ while the transfer of weight of the
${\rm Si}$ line is complete at 2 K), a precise structural determination of the
low temperature phase has not been possible yet. On the basis of the
results of the present work, we would like to propose that this is
related to the natural instability of the collinear phase toward
an orthorhombic distortion that we have identified in the $J_1{-}J_2$ model
coupled to the lattice. 

In summary, we have shown that frustration can lead to a Peierls-like
transition when the coupling to the lattice is taken into account.
This has been proven for the $J_1{-}J_2$ model, which is expected
to have a zero-temperature spontaneous instability towards
an orthorhombic phase for {\it infinitesimal spin-lattice coupling} when
the collinear order take place.
This effect, which is related to the presence of
a residual degeneracy of the GS after quantum fluctuations
have been taken into account, is expected to be efficient in other
quantum frustrated antiferromagnets, and work is in progress along
these lines. For the $J_1{-}J_2$ model, another surprise came from
the parameter range in which a dimerized phase is stabilized. This
phase seems to be limited to large values of the spin-lattice constant,
while it is also expected to occur for infinitesimal coupling if
dimer order is present in the gapped phase around $J_2/J_1=0.5$.
It would be very exciting to have a physical realization of the $J_1{-}J_2$
model in that parameter range to have an experimental input regarding
the relevance of that phase.

We would like to thank P. Carretta, S. Miyahara, D. Poilblanc, and
L. Capriotti for many useful discussions and the Swiss National Fund.


\end{document}